%% file: paper.tex
\documentclass[a4paper,fleqn]{cas-dc}
\usepackage{cas-dfrws}
\renewcommand{\dfrwsconference}{Digital Forensics Research Conference Europe (DFRWS EU) 2021, March 29--April 1, 2021}

\usepackage{graphicx}
\usepackage{tabularx}
\usepackage{listings}
\usepackage[numbers,sort]{natbib}

\usepackage[nolist]{acronym}

\usepackage{caption,subcaption}

\newcommand{\id}[1]{\textit{#1}}
\PassOptionsToPackage{hyphens}{url}
\usepackage{url}

\begin{document}
\input{acronyms}
\shorttitle{Bringing Forensic Readiness to Modern Computer Firmware}
\shortauthors{Latzo et~al.}
%shortauthors{Anonymous et~al.}

\title[mode = title]{Bringing Forensic Readiness to Modern Computer Firmware}

% DFRWS double blind review

%\author{Anonymous}
\author{Tobias Latzo\corref{mycorrespondingauthor}}
\ead{tobias.latzo@fau.de}
\credit{Conceptualization, Investigation, Methodology, Software, Supervision, Validation, Visualization, Writing -- original draft, Writing -- review \& editing}
\author{Florian Hantke}
\ead{florian.hantke@fau.de}
\credit{Conceptualization, Investigation, Software, Validation, Visualization, Writing -- original draft, Writing -- review \& editing}
\author{Lukas Kotschi}
\ead{lukas.kotschi@fau.de}
\credit{Conceptualization, Investigation, Software, Validation, Writing -- review \& editing}
\author{Felix Freiling}
\ead{felix.freiling@cs.fau.de}
\credit{Supervision, Methodology, Writing -- original draft, Writing -- review \& editing}
\address{Department of Computer Science, Friedrich-Alexander-Universit\"at Erlangen-N\"urnberg (FAU), Erlangen, Germany}
\cortext[mycorrespondingauthor]{Corresponding author}

\begin{abstract}
Today's computer systems come with a pre-installed tiny operating system, which is also known as UEFI. UEFI has slowly displaced the former legacy PC-BIOS while the main task has not changed: It is responsible for booting the actual operating system. However, features like the network stack make it also useful for other applications. This paper introduces UEberForensIcs, a UEFI application that makes it easy to acquire memory from the firmware, similar to the well-known cold boot attacks. There is even UEFI code called by the operating system during runtime, and we demonstrate how to utilize this for forensic purposes.
\end{abstract}


\maketitle

\section{Introduction}
The role of main memory analysis has become a more and more important part of digital forensic investigations. Main memory often contains valuable data that is never written to persistent storage. Examples are running processes, open network connections or even encryption keys that are necessary to decrypt containers. Furthermore, there is fileless malware that cannot be detected otherwise.

Memory acquisition has, therefore, become an essential part of forensic investigations. Ideally, memory acquisition cannot be detected by the target system, does not change data, and is performed atomically~\cite{voemelFreiling2012}. To provide the highest level of authenticity, memory acquisition should be performed on as low a level as possible, i.e., below the \ac{os} \cite{latzo2019universal}. However, main memory acquisition is much more intricate than acquiring hard disk contents. Nowadays, there are numerous ways of memory acquisition \cite{DBLP:journals/di/VomelF11}, e.g., using kernel support or modules like the Linux tools Pmem~\cite{pmem} or LiME~\cite{lime}. Virtualized target systems can often be halted and the memory acquired using built-in tools of the hypervisor. It is even possible to virtualize the target system on-the-fly~\cite{DBLP:conf/raid/MartignoniFPC10}. However, most of these techniques have the following requirements: (1) Forensic software has to be deployed on the target system at runtime, and (2) this software has to run with root privileges. Moreover, even if these requirements can be met, there are anti-forensic techniques that may circumvent or tamper with memory acquisition~\citep{shadowWalker, palutkeFreiling2018}. In general, it would therefore be useful if forensic software was ``pre-installed'' on the system, a condition known as \emph{forensic readiness} \cite{Rowlingson}.

The \ac{uefi} was introduced as the successor of the meanwhile nearly 40 years old PC-BIOS and is a ``pre-installed'' software, albeit not for forensic purposes. \ac{uefi} allows to start the \ac{os} in long mode, supports Secure Boot, and even own EFI applications can be executed in the \ac{uefi} Shell. Most \ac{uefi} implementations even come with a full network stack and sometimes even with a web browser. \ac{uefi} also specifies \acp{rts} that can be called by the \ac{os}. These allow, for example, reading and setting \ac{uefi} variables or updating the firmware. There is a \ac{uefi} reference implementation called \ac{edk2}~\cite{edk2}.

In this paper, we exploit modern computer firmware's high capabilities and bring forensic readiness to the \ac{uefi}. For this, we introduce
\begin{quote}
  \emph{\underline{UE}FI \underline{b}uilt-in m\underline{e}mo\underline{r}y \underline{forensics}}  
\end{quote}
(abbreviated as UEberForensIcs) for which we integrated forensic memory acquisition software. That can be used during the boot process. The memory acquisition is based on the concept of \emph{cold boot}, which is explained in more detail below. Furthermore, we show how to persist code in the \ac{uefi} \acp{rts} and get code execution that can also be used for forensic software. Additionally, we have built a tracer that traces calls of \ac{uefi} \acp{rts}. 
\subsection{Related Work}
\label{subsec:relatedWork}
In 2008, \citet{DBLP:conf/uss/HaldermanSHCPCFAF08} introduced \emph{cold boot} attacks. They exploited the fact that DRAM modules are not instantly cleared when unplugged, which is also known as the memory remanence effect. To acquire memory, they transplanted DRAM modules and attached it to an analysis system where the module is readout. During the replug procedure, the RAM module is cooled with coolant spray. Eventually, the researchers were able to restore encryption keys. 

While the focus of \citet{DBLP:conf/uss/HaldermanSHCPCFAF08} was key recovery after memory transplantation, the authors also mentioned the possibility of hard resets and booting a system for memory acquisition. In the latter, one has to deal with the BIOS footprint that overwrites few megabytes of the RAM. Further research~\cite{Gruhn2013} revealed that newer RAM modules are not as vulnerable to memory transplant attacks as the older DDR2 modules. Furthermore, newer modules do scramble data to avoid the parasitic effects of semiconductors. However, memory scrambling has not proven to be effective protection against cold boot attacks~\cite{bauer2016lest}.

Regarding related work on UEFI, we are aware of a Master's thesis~\cite{markanovic2014trusted}. The students made use of a signed UEFI application that was used to dump physical memory to a USB flash drive. They focused on building a static chain of trust whereby the trust anchor is the firmware. Furthermore, there is a blog post by Frisk who used the \ac{uefi} \acp{rts} to circumvent 4 GiB DMA limitations in PCILeech~\cite{ulfFrisk}. Usually, the Linux kernel is mapped into the upper physical memory. So PCILeech cannot inject code via 32 bit DMA. To get around this limitation, he manipulated the \ac{uefi} \acp{rts} function pointer table --- that is located in lower memory regions --- to inject own code.

ISO/IEC 27043:2015 \cite{ISO27043} defines \emph{digital forensic readiness} as the ``process of being prepared for a digital investigation
before an incident has occurred.'' Forensic readiness is related to preparation phases in many process models of incident response and digital forensic investigations and usually involves establishing a capability for securely gathering legally admissible evidence in case of an incident \cite{Rowlingson}. In practice, the quality of forensic readiness is closely related to the level of logging, the effectiveness of alerting and incident management processes and the quick availability of evidence acquisition capabilities which ideally are pre-deployed as software \cite{DBLP:journals/di/MoserC13} or in hardware \cite{DBLP:journals/di/CarrierG04}.
\subsection{Contribution}
This paper shows how to make a computer's \ac{uefi} forensic ready. The main contributions of this are as follows:
\begin{enumerate}
	\item We introduce UEberForensIcs show how to integrate forensic software that enables cold boot like memory acquisition directly into a computer's firmware. The evaluation in this paper reveals that this approach can also be practically used.
	\item Furthermore, we show how to persist code in the \ac{uefi} that is executed when the operating system is running. This code runs with kernel privileges and can also be used for memory acquisition.
	\item We developed an \ac{os}-independent \ac{rts} tracer. The \ac{rts} are thereby traced in the \ac{rts} itself. Our evaluation gives insights which and how often specific \ac{rts} are typically called in different scenarios.
\end{enumerate}
We have published UEberForensIcs~\footnote{\url{https://github.com/ueFAUrensics/UEberForensIcs}} and the RTS tracer~\footnote{\url{https://github.com/ueFAUrensics/RTStracer}} on Github.
\subsection{Outline}
In Section~\ref{sec:background} background information about the \ac{uefi} and \ac{edk2} is given. The architecture and setup of our experiments is described in Section~\ref{sec:architectureAndSetup}. Then, we show insights into the implementation of the built-in forensic acquisition software including an evaluation. In Section~\ref{sec:rtsForensics} we show how the \ac{rts} tracer runs with hooking. Finally, in Section~\ref{sec:conclusion} this paper is concluded.
\section{Background}
\label{sec:background}
In the following, we want to give some background information on some existing concepts that are used in this paper.
\subsection{Criteria for Memory Acquisition}
\label{subsec:critireaMemoryAcquisition}
\citet{voemelFreiling2012} defined three criteria for forensically sound memory acquisition: \emph{correctness}, \emph{atomicity} and \emph{integrity}. In the following, we want to briefly explain these criteria and show how to quantify them~\cite{gruhnFreiling2016}.

A memory snapshot is considered to be \emph{correct} if the used memory acquisition software acquired the memory's actual content. Obviously, this criteria is very fundamental for memory acquisition.

Memory acquisition software that is running on the target itself usually cannot stop all other system activity. This may lead to memory dumps that show the effects of events for which the actual cause has not been recorded. If such inconsistencies do not occur --- this is usually the case if the system can be halted --- a memory dump is called \emph{atomic}. \citet{gruhnFreiling2016} quantified the atomicity of a memory dump by the time between the acquisition of the first memory region and the last memory region.

\emph{Integrity} is ensured if the content of a memory image is not changed after an investigator decides to take a snapshot. \citet{voemelFreiling2012} state that integrity can be quantified by the level at which the process of taking the snapshot changes memory. 
\subsection{Unified Extensible Firmware Interface}
The \ac{uefi} was introduced in 1998 as a successor of the legacy PC-BIOS. Often the \ac{uefi} is still called BIOS. A more generic term for \ac{uefi} and BIOS is \emph{firmware}. The \ac{uefi} boots itself into protected mode (32 bit) or long mode (64 bit) instead of the real mode (16 bit). Obviously, this makes development much easier. As a consequence, the \ac{uefi} implementations are often tiny \ac{os}s with own applications, network stack, and so on. There are several specified stages, e.g., the Security (SEC) phase as the first stage, followed by the Pre EFI initialization (PEI) and \ac{dxe} phase. When reaching the \ac{dxe} phase, basically all hardware initializations happened, and hardware can be used.

In 2004, Intel released an open-source implementation called Tiano of an EFI. Tiano evolved to \ac{edk2} and is now maintained by the TianoCore community~\cite{edk2}.
\section{Architecture and Setup}	
\label{sec:architectureAndSetup}
Developing and debugging firmware is an intricate affair and usually requires a special setup to be performed. We now describe the setup in which we developed our system and performed the experiments.

\begin{figure*}[tbh]
\centering
\includegraphics[width=\textwidth]{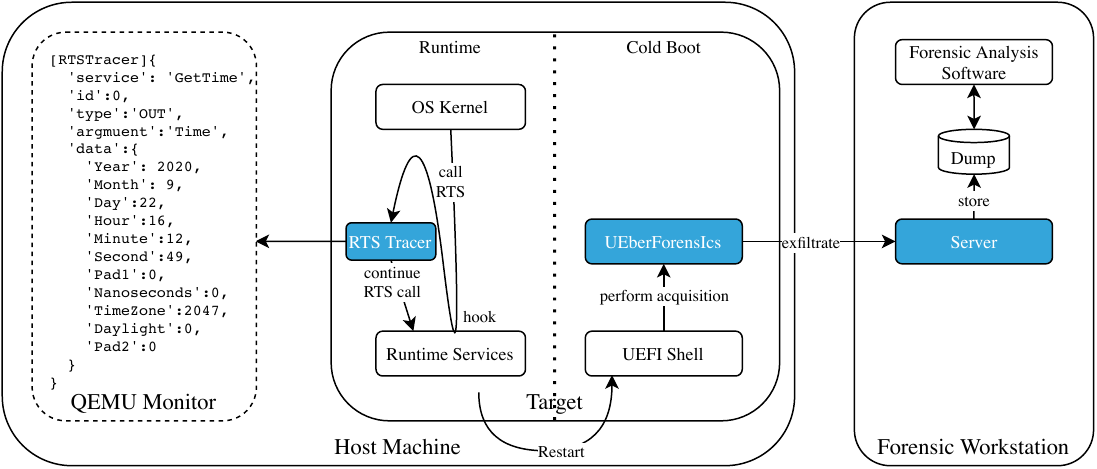}
\caption{Simplified architecture of UEberForensIcs and the RTS tracer.}
\label{fig:architecture}
\end{figure*}

Figure~\ref{fig:architecture} shows a simplified graph of the architecture we use for our experiments. Filled boxes indicate that these modules are our own developments. The target is running in a virtual machine with QEMU hypervisor. This makes development easier because, in this case, we do not need to reprogram SPI flash chips for every change. Furthermore, it simplifies debugging. The right side of the graph is dedicated to the built-in cold boot part (see also Section~\ref{sec:builtinColdBoot}) while the left side is dedicated to the runtime forensics part (see also Section~\ref{sec:rtsForensics}). 

\subsection{Hardware Setup}
 
The entire research was conducted on a standard laptop with an Intel Core i5-5200U CPU (2.20GHz, 2 cores) and 8 GiB of RAM. It runs Ubuntu Linux 18.04.4 with kernel version 4.15.0-118. The installed QEMU version is 4.2.92.

\subsection{VM Setup}
Usually, virtual machine monitors come with specialized firmware implementations. In most cases, emulating firmware is not intended. For virtual machines, there is a target for \ac{edk2} called OVMF. This port supports QEMU's virtual hardware. In Listing~\ref{lst:qemu} one can see the command to start the corresponding QEMU \ac{vm}. The \ac{vm} is running Ubuntu Linux 20.04 and has 2 GiB of RAM. 

\begin{lstlisting}[caption={Start of the virtual machine using QEMU.}, label={lst:qemu}, captionpos=b, breaklines=True, frame = single, basicstyle=\footnotesize, columns=flexible]
qemu -bios edk2/Build/OvmfX64/RELEASE_GCC5/FV/OVMF.fd
     -drive format=raw,file=ubuntu-linux.raw
     -drive format=raw,file=fat:rw:vm-content
     -global virtio-net-pci.romfile=""
     -nic tap,model=virtio-net-pci
     -m 2048M
     -debugcon file:debug.log
     -enable-kvm -cpu host
     -cdrom ubuntu-20.04-desktop-amd64.iso
\end{lstlisting}

Table~\ref{tab:ranges} shows the physical memory map of our virtual machine. In this case, memory is quite cohesive. Memory range \#3 is by far the largest memory region. A real system's RAM is usually more fragmented than those in QEMU. There is only a single memory hole from \#2 to \#3.
\begin{table*}[htb]
\centering
\caption{Memory map of the virtual machine we used for our experiments.}
\label{tab:ranges}
\begin{tabular}{rrrrrl}
\hline
\# & Start & End & Pages & Size & Purpose \\
\hline
1 & \texttt{0x00000000} & \texttt{0x0009ffff} & 160 & 640 kiB & System RAM\\
2 & \texttt{0x000a0000} & \texttt{0x000bffff} &  32 & 128 kiB & PCI Bus\\
3 & \texttt{0x00100000} & \texttt{0x7fffffff} & 524032 & 2047 MiB & System RAM\\
\hline	
\end{tabular}
\end{table*}
\section{Built-in Cold Boot}
\label{sec:builtinColdBoot}
In this section, we give insights into the implementation of UEberForensIcs. UEberForensIcs is a forensic cold boot like acquisition software that is integrated into the firmware. 

The use case of UEberForensIcs is that it is pre-installed on the firmware of a computer. While the \ac{os} is running, a potential incident happens, and so an incident responder is alerted. The incident responder wants to analyze what happened on the system with memory analysis. For the acquisition, they restart the computer into the EFI Shell and perform memory acquisition using UEberForensIcs. Therefore, the analyst needs no special equipment or installed tools on the host. The dump is transferred via network to the Forensic Workstation where it can be analyzed. 

\subsection{Implementation}
UEberForensIcs can be used as a standalone application or a dynamic command. For the evaluation, we used the latter variant. Basically, UEberForensIcs is implemented as a \ac{dxe} driver. 

We do not save the memory dump on the local drive because that would lead to corruption. Instead, we exfiltrate the data via the network (see also Figure~\ref{fig:architecture}). So UEberForensIcs requires an active network connection to the Forensic Workstation, and so we make use of \ac{edk2}'s TCP stack. The IP address is obtained via DHCP. When the connection is established, UEberForensIcs traverses the memory and sends it page-wise to the Forensic Workstation.
\subsection{Evaluation}
\label{subsec:builtin:eval}
\begin{figure*}[tbh]
 \centering
 \includegraphics[width=.9\textwidth]{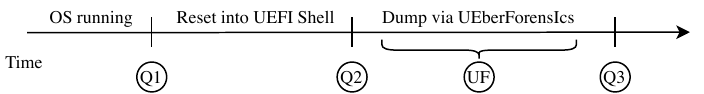}
 \caption{Timeline of the system with the four memory dumps.}
 \label{fig:timeline}
 \end{figure*}

\citet{gruhnFreiling2016} provided a framework to evaluate memory acquisition tools in terms of \emph{correctness}, \emph{atomicity} and \emph{integrity} \cite{voemelFreiling2012}. Basically, cold boot like attacks are performed atomically, i.e., the RAM module is removed, or there is a hard reset. So in this evaluation, we want to focus on correctness and integrity. For this purpose, we have generated the following four memory dumps, which one can see in Figure~\ref{fig:timeline}.
The dumps are acquired in the following way:

\textbf{Q1} The first dump is acquired using QEMU's \texttt{pmemsave} feature while the \ac{os} is running. Before \texttt{pmemsave} is started, the system is paused. We consider this dump as the \emph{ground truth}.

\textbf{Q2} The second dump is also performed using QEMU's \texttt{pmemsave} after the reset when the EFI Shell is started. To acquire memory atomically, the system is also paused. Furthermore, the \ac{os} is not running anymore. This means that processes that have run before do not alter memory anymore. However, \ac{edk2} also overwrites some smaller parts of memory.
  
\textbf{UF} The third dump is generated with our tool. Since the dump is performed on the same system, we cannot pause the system. However, we consider it to be atomic because the processes and any other thread of the \ac{os} processes are not running anymore. The only running activities belong to \ac{edk2} and so are not important.

\textbf{Q3} The last dump is acquired using QEMU's \texttt{pmemsave} after the UEberForensIcs dump is completed.

\begin{figure*}[htb]
%\begin{center}
\begin{subfigure}{0.3\textwidth}
	\centering
	\includegraphics[width=\textwidth]{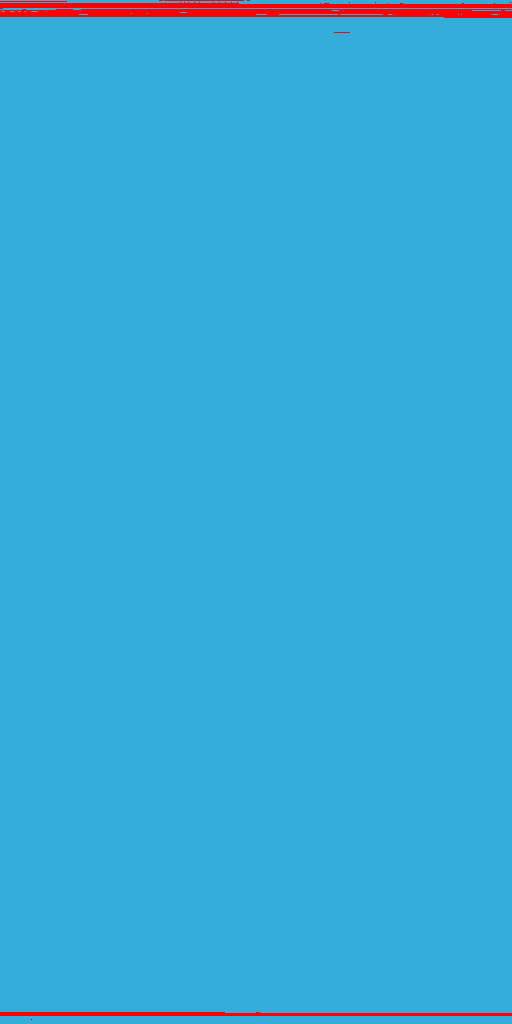}
	\caption{$\id{diff}(Q1, Q2)$: 24.6 MiB}
	\label{subfig:q1q2}
\end{subfigure}\hfil
\begin{subfigure}{0.3\textwidth}
	\centering
	\includegraphics[width=\textwidth]{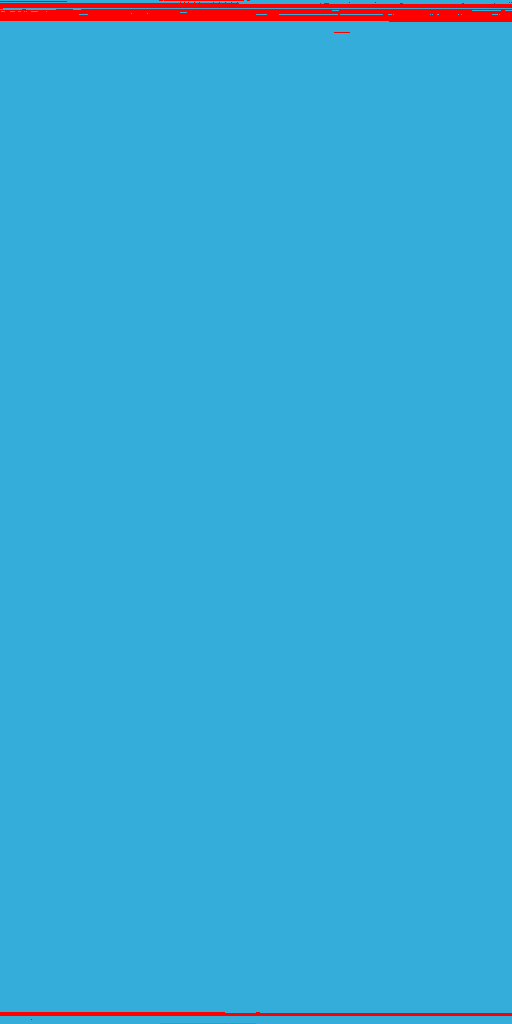}
	\caption{{$\id{diff}(Q1, \id{UF})$)}: 29.1 MiB}
	\label{subfig:q1uf}

\end{subfigure}\hfil
\begin{subfigure}{0.3\textwidth}
	\centering
	\includegraphics[width=\textwidth]{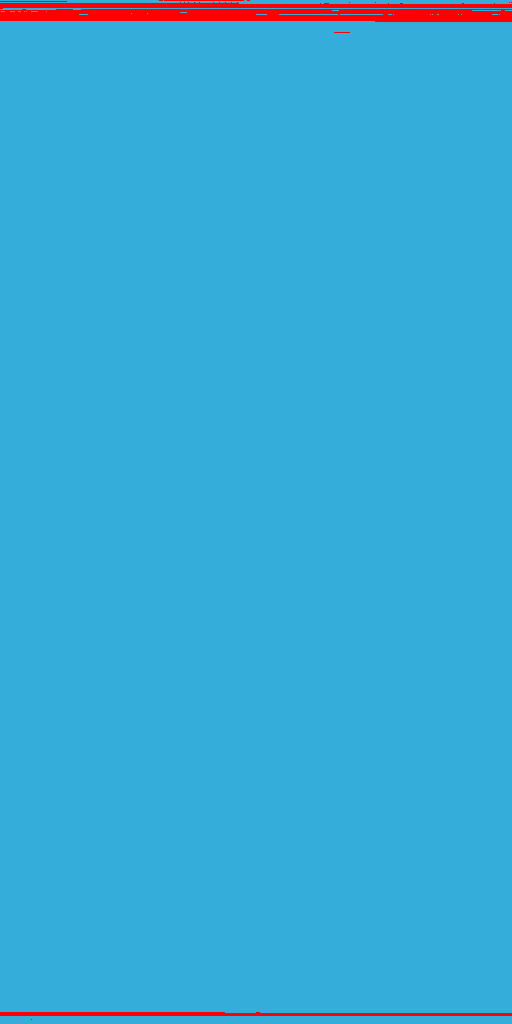}
	\caption{$\id{diff}(Q1, Q3$): 32.7 MIB}
	\label{subfig:q1q3}
\end{subfigure}\hfil

\medskip

\begin{subfigure}{0.3\textwidth}
	\centering
	\includegraphics[width=\textwidth]{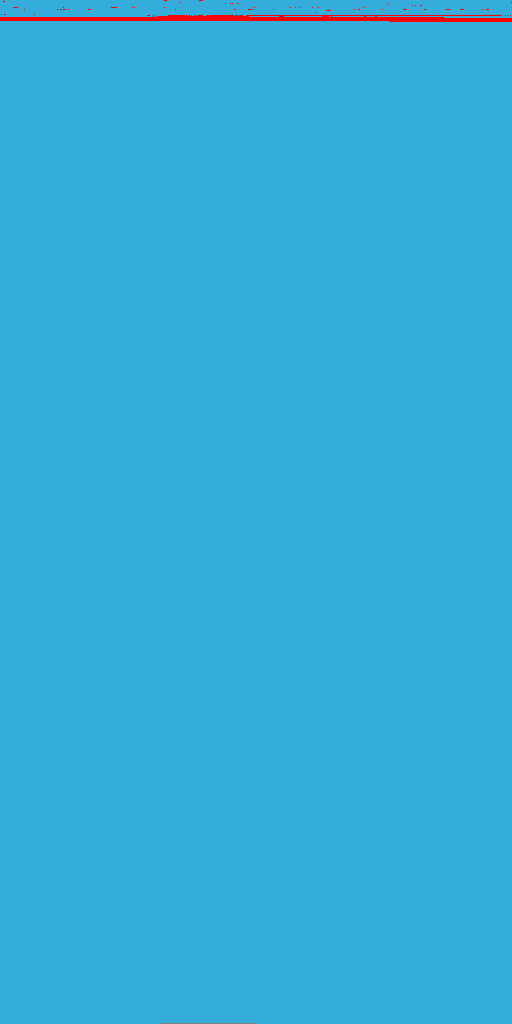}
	\caption{$\id{diff}(Q2, \id{UF})$): 4.9 MiB}
	\label{subfig:q2uf}

\end{subfigure}\hfil
\begin{subfigure}{0.3\textwidth}
	\centering
	\includegraphics[width=\textwidth]{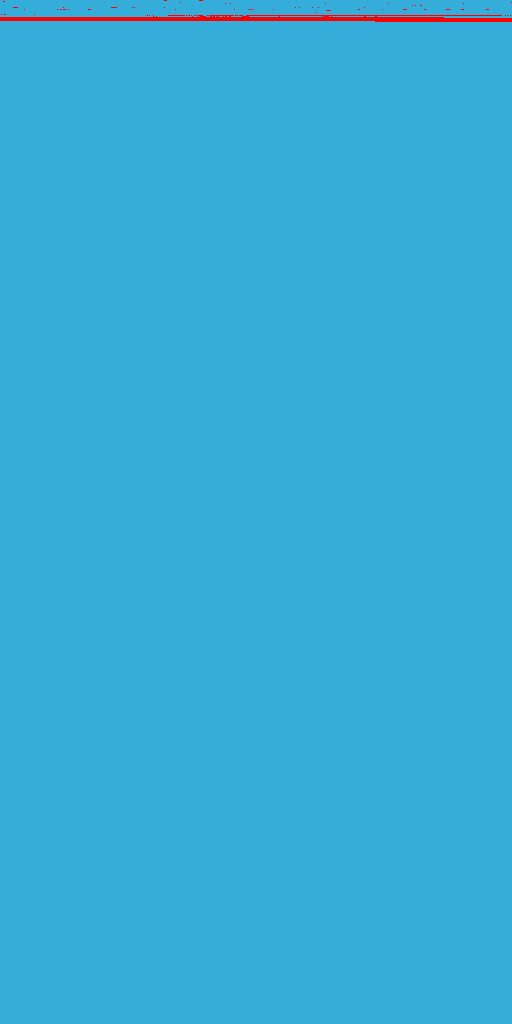}
	\caption{{$\id{diff}(Q2, Q3$)}: 8.5 MiB}
	\label{subfig:q2q3}

\end{subfigure}\hfil
\begin{subfigure}{0.3\textwidth}
	\centering
	\includegraphics[width=\textwidth]{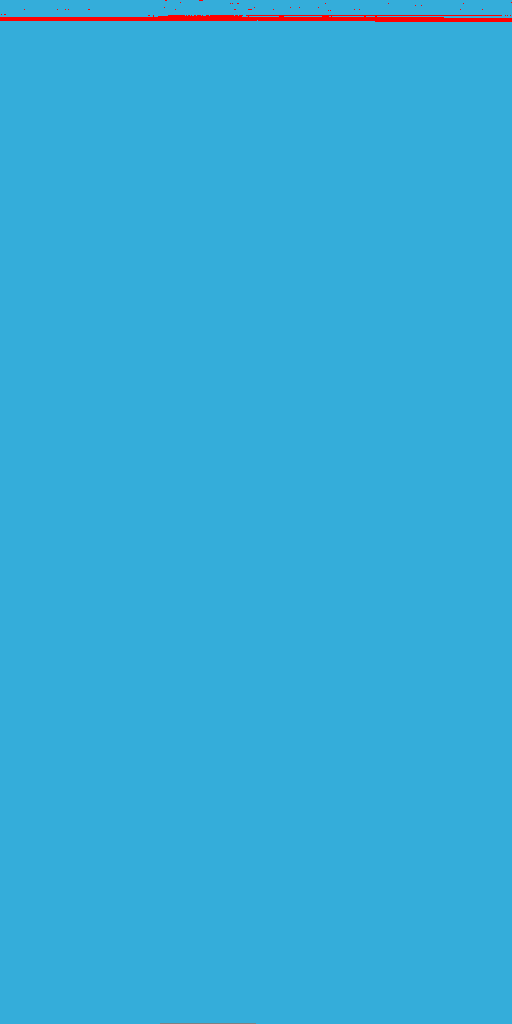}
	\caption{{$\id{diff}(\id{UF}, Q3$)}: 5.8 MiB}
	\label{subfig:ufq3}

\end{subfigure}\hfil
%\end{center}
\caption{Visualization of the page-wise diff. Addresses are growing from the left to the right and from the bottom to the top.}
\label{fig:pwdiff}
\end{figure*}

The evaluation of correctness and integrity is based on the analysis of differing bytes and pages of different dumps. In Table~\ref{tab:ueberforensics:eval} one can see the results of pairwise dumps. A visualization of page-wise (4 kiB) diffs can be found in Figure~\ref{fig:pwdiff}. A blue pixel indicates that the corresponding page is the same in both dumps. A red pixel indicates that the corresponding pages are differing by at least one byte. There are 1024 rows with 512 pages per line, i.e., 2 MiB per line. Addresses are growing from left to right and from the bottom to the top.

In the following two sections, we use these results to show to what extent UEberForensIcs affects memory and argue that UEberForensIcs works properly.
\begin{table*}[htb]
\centering
\caption{The table shows the results of differing bytes of the dumps and the corresponding proportion of total memory that is changed.}
\label{tab:ueberforensics:eval}
\begin{tabular}{rccrrrr}
\hline
\# & Dump 1 & Dump 2 & Total Pages & Total Size & Proportion \\
\hline
1 & Q1 & Q2 & 8143 & 24.6 MiB & 1.2 \%\\
2 & Q1 & UF & 10245 & 29.7 MiB & 1.4 \%\\
3 & Q1 & Q3 & 10260 & 32.7 MiB & 1.6 \%\\
4 & Q2 & UF & 2634 & 4.9 MiB & 0.2 \%\\
5 & Q2 & Q3 & 2568 & 8.5 MiB & 0.4 \%\\
6 & UF & Q3 & 2588 & 5.8 MiB & 0.3 \%\\ 
\hline
\end{tabular}
\end{table*}
\subsubsection{Correctness}
First, we want to show that UEberForensIcs works properly. To show this, we compare the ground truth (Q1) with the UEberForensIcs dump (UF), i.e., $\id{diff}(Q1, \id{UF})$ and with the dump Q3, i.e., $\id{diff}(Q1, Q3)$. It is striking that the total differing numbers are in the same order of magnitude (see also Table~\ref{tab:ueberforensics:eval}). Furthermore, the corresponding diffs' visualizations in  Figure~\ref{subfig:q1uf} and Figure~\ref{subfig:q1q3} show that the diffs are very similar. The ranges of diffing memory regions are basically the same for all dumps. Note that the dump of UF is made sequentially and transferred via network. The dump of Q3 is acquired atomically after the execution of UEberForensIcs. So, these diffs are not completely the same.

The $\id{diff}(Q2, \id{UF})$ shows that the QEMU \texttt{pmemsave} dump and the UEberForensIcs dump are differing in about 5 MiB. The corresponding visualization in Figure~\ref{subfig:q2uf} also reveals that the corresponding memory regions are basically the same. For $\id{diff}(\id{UF}, Q3)$ the diff is around 5.8 MiB. However, as Figure~\ref{subfig:ufq3} shows, the differing pages are in the upper memory regions as before.

The comparisons of the diffs showed that the dump of UEberForensIcs looks reasonable. Basically, the only differing pages are located in the upper memory regions that we can also observe with the QEMU built-in \texttt{pmemsave}.  

\subsubsection{Integrity}
Memory acquisition using UEberForensIcs is performed on the target system. This means that we do change memory. In this section, we show how much memory is changed. We also show which parts of the memory get changed by UEberForensIcs.

Figure~\ref{fig:pwdiff} gives a good impression of what and how much memory is changed when using UEberForensIcs. All diffs with the dump when the OS was running (Q1) show that some memory is overwritten in the lower memory regions --- basically starting at 0x1000000 --- when the computer is reset. This region has a size of about 7 MiB. When the system is restarted, there is no change in this memory region anymore (see also Figure~\ref{subfig:q2uf}, Figure~\ref{subfig:q2q3} and Figure~\ref{subfig:ufq3}).

Furthermore, Figure~\ref{fig:pwdiff} also shows that the reboot of the system has the most impact. The execution of UEberForensIcs also has impact (see also Figure~\ref{subfig:q2uf}, Figure~\ref{subfig:q2q3} and Figure~\ref{subfig:ufq3}). However, most of these memory regions are changed because of the reboot anyway. 

Overall, we can say that the execution of UEberForensIcs changes about 32 MiB of the whole memory. Thereof the most considerable amount is overwritten because of the reboot that loads the firmware. The majority of the firmware in our environment was located in the upper memory regions. There was no single differing byte in the middle of the memory.
\subsection{Discussion}
Writing software for \ac{uefi} is much easier than for the former PC-BIOS. \ac{edk2} code is written in C, and there are many features like a full network stack that facilitate the development of their own software. 

The evaluation showed that UEberForensIcs is working correctly. However, the whole acquisition process using UEberForensIcs changes about 30 MiB of RAM. This may differ from setup to setup depending on the firmware's footprint. In our setup, the memory ranges that were changed are located on the upper and lower border of the RAM, and so we argue that memory acquisition using UEberForensIcs is practical since we can acquire most memory atomically and do not rely on software on the host that may be manipulated by malware. 

However, there are also countermeasures for cold boot attacks. RAM reset on reboot~\cite{Gruhn2013}, memory scrambling~\cite{bauer2016lest} or locking the firmware that an adversary cannot boot from an own device are three examples. In our scenario, we control the firmware. So we can control that such countermeasures are not implemented or are only effective during normal reboots and not when an analyst is present. A particular hardware device could indicate this. 
\section{Runtime Service Forensics}
\label{sec:rtsForensics}
In the previous section, we have seen how our UEFI driver can be used to perform cold boot attacks. Now, we want to provide the first steps towards forensic memory acquisition using UEFI drivers at runtime. Thus, incident response teams could extract memory without rebooting and installing any memory acquisition software on the target that would change evidence.

In the following sections, we describe how to persist forensic tools in the \ac{uefi} \acp{rts} and gain code execution. For this, we provide a proof-of-concept tool that traces all \ac{rts} calls. Basically, this technique can also be used to perform memory acquisition in the \acp{rts}.
\subsection{Implementation}
Same as the former tool, the runtime tool is developed as a \ac{dxe} driver. However, this time the driver needs to continue execution even after \texttt{ExitBootServices()} is called, something common drivers do not fulfill. Therefore, the driver needs to be of the type \texttt{DXE\_RUNTIME\_DRIVER}.

Runtime drivers start in the \ac{dxe} phase and continue executing after the boot process is finished when the \ac{os} is running. Also, they have access to both \ac{rts} and boot services. Boot services stop to work after the OS loader calls \texttt{ExitBootServices()}, \acp{rts} persist after the \ac{dxe} phase. Their pointers get converted from physical addresses to virtual ones when \texttt{SetVirtualAddressMap()} is called.

To execute code after the \ac{dxe} phase, the driver needs to be called by another instance, such as the \ac{os}. The \ac{os} in our research is an instance we do not control. Hence we decided to take \acp{rts} that are called by the \ac{os} as a trigger for code execution. To activate the code execution, we implemented hooks for all \acp{rts} and set them in the \acp{dxe} phase when it initializes our driver. Therefore, the driver stores the origin service pointer and replaces its address table entry with a pointer to our hook. Furthermore, it registers a notifier to react when the \ac{os} calls \texttt{SetVirtualAddressMap()} and converts all pointers.

The hooks allow us to execute arbitrary code at runtime, which we developed further to implement an open-source \ac{rts} tracer. With the tracer, we can follow the called services and view their arguments to analyze the UEFI behavior thoroughly. The tracer outputs its information in JSON format.

An example call can be seen in Listing~\ref{lst:rtsLog}. Every JSON object contains one argument of the called service and its data. The JSON is limited to a maximum of 255 characters, which is why not all arguments fit in one object. Additionally, every argument is either of the type INput or OUTput and accordantly listed before or after the origin call. The example shows two output arguments of the service \textit{GetTime}.
\begin{lstlisting}[float, caption={The log shows an example result provided by the RTS tracer.}, label={lst:rtsLog}, captionpos=b, frame=single]
[RTSTracer]{
  'service': 'GetTime',
  'id':0,
  'type':'OUT',
  'argmuent':'Time',
  'data':{
    'Year': 2020,
    'Month': 9,
    'Day':22,
    'Hour':16,
    'Minute':12,
    'Second':49,
    'Pad1':0,
    'Nanoseconds':0,
    'TimeZone':2047,
    'Daylight':0,
    'Pad2':0
  }
}
[RTSTracer]{
  'service': 'GetTime', 
  'id':0, 
  'type':'OUT', 
  'argument':'Capabilities',
  'data':{
    'Resolution': 0, 
    'Accuracy': 0, 
    'SetsToZero':0
  }
}
\end{lstlisting}

\subsection{Evaluation}
The evaluation of the runtime service section is split into two parts. First, we show that we have arbitrary code execution at any time at runtime. Second, we evaluate the data created by the \ac{rts} tracer and compare various scenarios.

As mentioned before, we make use of runtime service hooks to execute code in UEFI. UEFI code execution at runtime could be used by incident response teams to extract memory without rebooting the \ac{os}. A requirement for this is that it can be executed at any time. However, our trigger depends on \acp{rts} being called, which is not often the case after the user logged in. Nevertheless, our tests show that the OS calls the \ac{rts} \texttt{GetVariable} whenever the user reads the \textit{efivars} (\texttt{/sys/firmware/efi/efivars/}). To prove this, we successfully modified our hook to force \texttt{System\_Reset} as soon as we read \textit{efivars}. Thus we fulfill the requirement to execute code at any time at runtime, which finishes the first part of the evaluation.

For the second part, we evaluate the \ac{rts} tracer. Therefore, we recorded the \ac{rts} calls on our Ubuntu VM in six different scenarios:
\begin{itemize}
	\item \textit{Boot} - We started the VM but did not log in.
	\item \textit{Login} - We started the VM and logged in as the user.
	\item \textit{Working} - We started the VM, logged in as the user, and performed standard working tasks for 15 minutes. These tasks were reading, writing, and configuring OS settings.
	\item \textit{Hour} - We started the VM, logged in as the user, and let it run for one hour. The power save mode caused a lock screen, which we unlocked in the end.
	\item \textit{Switch} - We started the VM and logged in as the user. Afterward, we switched the user.
	\item \textit{Reboot} - We started the VM, logged in as the user, and rebooted the machine. Then we logged in again.
\end{itemize}
For every scenario, our \ac{rts} tracer collected data. We wrote a parser in Python which is also included in the RTS tracer Github, to analyze and interpret the information. Table~\ref{tab:rtsCalls} shows the summarized results. It shows how often which runtime service was called in each scenario.
\begin{table*}[htb]
	\centering
	\caption{The table shows the number of RTS calls in various scenarios}
	\begin{tabular}{lrrrrrr } 
	\hline
 	Runtime Service  	& Boot & Login	& Working	& Hour	& Switch 	& Reboot 	\\ \hline 
 	\texttt{GetTime} 	& 46 & 46   	& 46 		& 46	 	& 46		& 92		\\ 
 	\texttt{GetVariable} 	& 754 & 786   	& 786 	& 786 	& 850	& 1617	\\ 
 	\texttt{SetVariable} 	& 110 & 110   	& 110	& 110 	& 110	& 165	\\ 
 	\texttt{GetNextVariableName} & 499 & 499   	& 499	& 499 	& 499	& 1067	\\ 
 	\texttt{ConvertPointer} & 91 & 91   	& 91 		& 91	 	& 91		& 182	\\ \hline 
 	Total 			& 1500 & 1532   	& 1532 	& 1532 	& 1596	& 3123	\\  
 	\hline
	\end{tabular} 
  	\label{tab:rtsCalls}
\end{table*}
The table lists only five of the 14 runtime services that are available according to section 5 in the \ac{edk2} UEFI Driver Writer's Guide~\cite{edk2DriverGuide}. Even if the guide lists more services, we only observed these five services in all scenarios.

When we look at the different scenarios, we can see that \emph{Login}, \emph{Working}, and \emph{Hour} have the same number of services called. Further, going into more detail, we can see that the calls' arguments are the same every time. This means that the \acp{rts} used during the startup routine remain the same. Moreover, even without studying the \ac{edk2} source, we can conclude from the same call number in the three scenarios that during standard OS usage, no \ac{rts} is used after the login.

The login and logout processes, on the other hand, make use of \acp{rts}. This is shown by the difference of counted calls in the scenarios \emph{Boot}, \emph{Login}, and \emph{Switch}. We see that 32 additional calls are registered on login and 32 more on logout in the results. All of them are \texttt{GetVariable} calls that request either the \texttt{OsIndicationsSupported} or the \texttt{OsIndications} variable. Both variables tell the OS which UEFI firmware features are supported and activated.

The last scenario, \emph{Reboot}, is different from the previous ones as the counted call number is a lot more. For \emph{GetTime} and \emph{ConvertPointer} the numbers are twice as large compared to the \emph{Login} scenario. This makes sense as we boot the system two times. On the other hand, we count 45 more \textit{GetVariable} calls, 69 more \textit{GetNextVariableName} calls, and 55 less \textit{SetVariable} calls on the second boot process. This is because the second boot process does not register every variable again but uses initialized variables from the first boot process. The variable \textit{OsIndications}, for instance, is only set in the first boot process. Afterward, it is requested 45 times during the first boot and 46 times during the second boot.

As shown above, the RTS tracer works well and gives clear insights into the usage of \acp{rts}.
% The use of the services is generally constant and stops after a successful login.
%
\subsection{Discussion}
\label{subsec:rts:discussion}
In this section, we showed how to gain code execution from the \ac{uefi} during \ac{os} runtime. As a proof-of-concept, we implemented an \ac{rts} tracer. The corresponding code is not resident on the hard drive but on the SPI flash chip and copied to the RAM by the system's firmware. However, in contrast to \ac{smm}-based approaches~\cite{crashsmm}, \acp{rts} are not executed on a higher privilege level but on the same as the \ac{os}. Developing code for the \ac{rts} is much easier than SMM's 16-bit Real Mode code. 

It is also possible to perform memory acquisition from the \acp{rts}. However, the exfiltration of memory is more complicated than in UEberForensIcs. The \ac{os} manages the network stack and has configured the network interface card. Other possibilities are to use persistent storage. Nevertheless, similar to the network interface, the \ac{os} manages hard drives. So it is not easy to use the \ac{rts} for memory exfiltration without adapting the \ac{os} kernel or drivers. 
\section{Conclusion and Future Work}
\label{sec:conclusion}
In this paper, we introduced UEberForensIcs that brings forensic memory acquisition to modern computer firmware. With UEberForensIcs, an analyst can perform simple cold boot attacks without any craftsmanship. Additionally, the dump can be considered to be atomic. The only precondition is that the UEFI is forensic-ready, i.e., UEberForensIcs must be part of the UEFI before the need for memory acquisition arises.

Our evaluation showed that only small distinct parts of the memory get overwritten by the firmware. For the development and evaluation, we used a QEMU \ac{vm}. However, future work should also consider compatibility and other possible memory layouts on actual physical systems.

Furthermore, we demonstrated how to gain code execution with kernel privileges without injecting code into the kernel without any persistent file on the hard drive. Therefore, we hook the \ac{uefi} \acp{rts}. As a proof-of-concept, we developed an \ac{rts} tracer that traces all occurring \ac{rts} calls of the OS kernel. The discussion in Section~\ref{subsec:rts:discussion} yielded that integrating memory acquisition software in the \acp{rts} can be beneficial. However, it is hard to exfiltrate data from there. 
\section*{Acknowledgments}
This research is supported by Deutsche Forschungsgemeinschaft (DFG, German Research Foundation) as part of the Research and Training Group 2475 "Cybercrime and Forensic Computing" (grant number 393541319/GRK2475/1-2019).

%
% author contribution according to https://casrai.org/credit/
\printcredits

\bibliography{literature}

\end{document}

%% file: acronyms.tex
\begin{acronym}
	\acro{dxe}[DXE]{\emph{Driver Execution Environment}}
	\acro{edk2}[EDK II]{\emph{EFI Development Kit II}}
	\acro{uefi}[UEFI]{\emph{Unified Extensible Firmware}}
	\acro{rts}[RTS]{\emph{Runtime Service}}
	\acroplural{rts}[RTSs]{Runtime Services}
	\acro{os}[OS]{operating system}
	\acro{smm}[SMM]{System Management Mode}
	\acro{tool}[UEberForensIcs]{\emph{\underline{UE}FI \underline{b}uilt-in m\underline{e}mo\underline{r}y \underline{forensics}}}
	\acro{vm}[VM]{\emph{Virtual Machine}}
\end{acronym}